\documentclass[aps,prd,twocolumn,nofootinbib,superscriptaddress]{revtex4}
\usepackage{amsmath,graphicx}
\usepackage{subfigure}

\usepackage[dvipdfm,colorlinks=true,citecolor=blue,pdfstartview=FitH]{hyperref}

\usepackage{color,framed} 

\def\bea{\begin{equation}}
\def\eea{\end{equation}}

\newcommand{\rt}{Regge trajectory}
\newcommand{\rts}{Regge trajectories}
\newcommand{\tr}{trajectory}
\newcommand{\trs}{trajectories}
\newcommand{\bfr}{{\bf r}}
\newcommand{\bfp}{{\bf p}}
\newcommand{\bfpa}{{|\bf p|}}
\newcommand{\gev}{{\rm GeV}}

\newcommand{\cltb}{$\bar{3}_c$}
\newcommand{\cltba}{\bar{3}_c}

\begin{document}
\title{$\lambda$ and $\rho$ Regge trajectories for the pentaquark $P_{cc\bar{c}bb}$ in the diquark-triquark picture}
\author{He Song}
\email{songhe\_22@163.com}
\affiliation{School of Physics and Information Engineering, Shanxi Normal University, Taiyuan 030031, China}
\author{Xin-Ru Liu}
\email{1170394732@qq.com}
\affiliation{School of Physics and Information Engineering, Shanxi Normal University, Taiyuan 030031, China}
\author{Jia-Qi Xie}
\email{1462718751@qq.com}
\affiliation{School of Physics and Information Engineering, Shanxi Normal University, Taiyuan 030031, China}
\author{Jiao-Kai Chen}
\email{chenjk@sxnu.edu.cn, chenjkphy@outlook.com (corresponding author)}
\affiliation{School of Physics and Information Engineering, Shanxi Normal University, Taiyuan 030031, China}

\begin{abstract}
We propose the Regge trajectory relations for the fully heavy pentaquark $P_{cc\bar{c}bb}$ utilizing both diquark and triquark Regge trajectory relations. Using these new relations, we discuss four series of Regge trajectories: the $\rho_1$-, $\rho_2$-, $\lambda_1$-, and $\lambda_2$-trajectories. We provide rough estimates for the masses of the $\rho_1$-, $\rho_2$-, $\lambda_1$-, and $\lambda_2$-excited states.
Except for the $\lambda_1$-trajectories, the complete forms of the other three series of Regge trajectories for the pentaquark $P_{cc\bar{c}bb}$ are lengthy and cumbersome. We show that the $\rho_1$-, $\rho_2$-, and $\lambda_2$-trajectories can not be obtained by simply imitating the meson Regge trajectories because mesons have no substructures. To derive these trajectories, pentaquark's structure and substructure should be taken into consideration. Otherwise, the $\rho_1$-, $\rho_2$-, and $\lambda_2$-trajectories must rely solely on fitting existing theoretical or future experimental data. Consequently, the fundamental relationship between the slopes of the obtained trajectories and constituents' masses and string tension will become unobvious, and the predictive power of the Regge trajectories would be compromised.
Moreover, we show that the lengthy complete forms of the $\rho_1$-, $\rho_2$-, and $\lambda_2$-trajectories can be well approximated by the simple fitted formulas. Four series of Regge trajectories for the pentaquark $P_{cc\bar{c}bb}$ all exhibit a behavior of $M{\sim}x^{2/3}$, where  $x=n_{r_1},n_{r_2},l_1,l_2,N_{r_1},N_{r_2},L_1,L_2$.
All four series of trajectories exhibit concave downward behavior in the $(M^2,\,x)$ plane.
\end{abstract}

\maketitle


\section{Introduction}
As a type of exotic hadrons, pentaquarks will enhance the study of hadrons and will provide new probes for understanding QCD \cite{LHCb:2015yax,
ParticleDataGroup:2024cfk,Jaffe:2004ph,Ali:2017jda,Liu:2019zoy,Karliner:2003dt,
Roca:2015dva,Lebed:2015tna,Okiharu:2004wy,Lebed:2016hpi,Olsen:2017bmm,
Brambilla:2019esw}.
In 2015, the first observation of pentaquark was reported by the LHCb collaboration in $\Lambda^0_b\to J/{\psi}K^-p$ decays \cite{LHCb:2015yax}.
In recent years, exotica states containing fully heavy quarks have been observed.
In 2020, a narrow structure around 6.9 ${\rm{GeV}/c^2}$ matching the lineshape of a resonance and a broad structure just above twice the $J/\psi$ mass were observed by the LHCb Collaboration \cite{LHCb:2020bwg}.
In 2023, the ATLAS Collaboration observed an excess of di-charmonium events in
the four-muon final state with the ATLAS detector \cite{ATLAS:2023bft}.
In 2023, the CMS Collaboration reported new structures in the $J/{\psi}J/\psi$ mass spectrum in Proton-Proton Collisions at $\sqrt{s} = 13\, {\rm TeV}$ \cite{CMS:2023owd}.

These observations suggest the existence of fully heavy pentaquarks and motivate the studies on the fully heavy pentaquarks.
Theoretical approaches include the quark model \cite{Gordillo:2024blx,Liang:2024met,Gordillo:2023tnz}, chromomagnetic interaction model \cite{An:2020jix,An:2022fvs}, extended Gursey-Radicati formalism \cite{Sharma:2025adr}, quark delocalization color screening model \cite{Yan:2021glh}, MIT bag model \cite{Zhang:2023hmg}, sum rules \cite{Wang:2021xao,Azizi:2025fmx,Zhang:2020vpz}, lattice-QCD inspired quark model \cite{Yang:2022bfu}, and the effective quark
mass and screened charge scheme \cite{Rashmi:2024ako}, among others.

The {\rt}\footnote{A {\rt} of bound states is generally expressed as $M=m_R+\beta_x(x+c_0)^{\nu}$ $(x=l,\,n_r)$ \cite{Chen:2022flh,Chen:2021kfw}, where $M$ is the mass of the bound state, $l$ is the orbital angular momentum, and $n_r$ is the radial quantum number. $m_R$ and $\beta_x$ are parameters. For simplicity, plots in the $(M,\,x)$ plane \cite{Xie:2024lfo}, $(M-m_R,\,x)$ plane \cite{Chen:2023cws}, $(M,\,(x+c_0)^{\nu})$ plane \cite{Burns:2010qq}, $(M^2,\,x)$ plane \cite{Chen:2018nnr}, $((M-m_R)^2,\,x)$ plane \cite{Chen:2023djq,Chen:2023web} or $((M-m_R)^{1/{\nu}},\,x)$ plane \cite{Xie:2024dfe}, are all commonly referred as Chew-Frautschi plots. {\rts} can be plotted in these various planes. } is one of the effective approaches widely used in the study of hadron spectra
\cite{Burns:2010qq,Regge:1959mz,Chew:1962eu,Nambu:1978bd,Gross:2022hyw,Brodsky:2006uq,Nielsen:2018uyn,
brau:04bs,Brisudova:1999ut,Guo:2008he,Ebert:2009ub,Irving:1977ea,Collins:1971ff,
Inopin:1999nf,Afonin:2014nya,MartinContreras:2020cyg,Sergeenko:1994ck,Veseli:1996gy,
Inopin:2001ub,Wilczek:2004im,
Sonnenschein:2018fph,MartinContreras:2023oqs,Roper:2024ovj,G:2024zkc,Oudichhya:2024ikt}. 
Few studies have applied the {\rt} approach to the discussion of pentaquarks. In Ref. \cite{Ghosh:2017cck}, {\rts} for the pentaquarks containing one heavy quark were investigated. In Ref. \cite{Sindhu:2023oqo}, {\rts} for the pentaquarks containing two charm quarks were discussed. In both of these references, only one series of {\rts} are discussed.
To our knowledge, there have been no systematic studies up to now addressing both two series of $\lambda$-{\trs} and two series of $\rho$-{\trs} for pentaquarks.
In the diquark-triquark picture, a pentaquark consists of a diquark and a triquark, where the triquark is composed of an antiquark and a diquark (see Fig. \ref{fig:pr}). By employing the diquark {\rt} relations \cite{Feng:2023txx} and the triquark {\rt} relations \cite{Song:2024bkj}, we attempt to apply the {\rt} approach to the pentaquark $P_{cc\bar{c}bb}$.
The masses of two $\lambda$-mode and two $\rho$-mode excited states are roughly estimated. Four series of {\rts}--the $\lambda_1$-, $\lambda_2$-, $\rho_1$-, and $\rho_2$-trajectories--are investigated.

The paper is organized as follows: In Sec. \ref{sec:rgr}, the {\rt} relations for the fully heavy pentaquarks are proposed. In Sec. \ref{sec:rts}, four series of masses and four series of {\rts} are presented. In Sec. \ref{sec:dis}, the discussions are given. The conclusions are presented in Sec. \ref{sec:conc}.

\section{{\rt} relations for the fully heavy pentaquarks}\label{sec:rgr}
In this section, by utilizing the diquark {\rts} \cite{Feng:2023txx} and the triquark {\rts} \cite{Song:2024bkj}, we propose the pentaquark {\rts}, which can be employed to discuss both $\lambda$- and $\rho$-trajectories.

\subsection{Preliminary}\label{subsec:prelim}

\begin{figure}[!phtb]
\centering
\includegraphics[width=0.26\textheight]{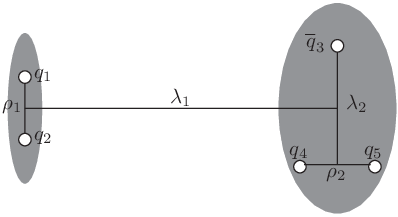}
\caption{Schematic diagram of a pentaquark in the diquark-triquark picture. The left grey part represents the diquark $1$, composed of two quarks ($q_1$ and $q_2$). The right grey part represents a triquark composed of one antiquark ($\bar{q}_3$) and diquark $2$. Diquark $2$ is composed of quarks $q_4$ and $q_5$. The circles denote the quarks and the antiquark.}\label{fig:pr}
\end{figure}

In the diquark-triquark picture, a pentaquark consists of one diquark and one triquark (see Fig. \ref{fig:pr}). $\rho_1$ and $\rho_2$ separates the quarks in the diquark $1$ and $2$, respectively. $\lambda_1$ corresponds to the separation between diquark $1$ and the triquark while $\lambda_2$ corresponds to the separation between the antiquark $\bar{q}_3$ and diquark $2$. There exist four excited modes: the $\rho_1$-mode involves the radial and orbital excitation in the diquark $1$; the $\rho_2$-mode involves excitation in the diquark $2$; the $\lambda_1$-mode involves the radial or orbital excitation between the diquark $1$ and triquark; and the $\lambda_2$-mode involves excitation between the antiquark and diquark $2$. Consequently, there are four series of {\rts}: two series of $\rho$-{\rts} and two series of $\lambda$-{\rts}.

A diquark $(q_1q_2)$ can couple to only two irreducible color representations: $3_c\otimes3_c=\cltba\oplus{6}_c$. The $\bar{3}_c$ is the attractive channel, while in the $6_c$ representation, the internal interaction between the $q_1q_2$ pair is repulsive.
In some works, both $\cltba$ and ${6}_c$ are considered, for example, see Refs. \cite{Maiani:2019lpu,Berwein:2024ztx}, while in other works, only $\cltba$ is considered, for example, see Refs. \cite{Brodsky:2014xia,Galkin:2023wox}.
Following Refs. \cite{Brodsky:2014xia,Galkin:2023wox}, only the $\bar{3}_c$ diquark is considered in the present work. Similarly, the triquark under consideration, $(\bar{q}_3(q_4q_5))$, is a ${3}_c$ bound state consisting of a diquark $(q_4q_5)$ in the $\cltba$ representation and an antiquark $\bar{q}_3$ also in the $\cltba$ representation. In the triquark-diquark model, the color-singlet pentaquarks under consideration are composed of a diquark in $\bar{3}_c$ and a triquark in $3_c$ \cite{Lebed:2015tna}.

In the diquark-triquark picture, the state of the pentaquark is denoted as
\bea\label{tetnot}
\left((q_1q_2)^{{\bar{3}_c}}_{n_1^{2s_1+1}l_{1j_1}}
\left(\bar{q}_3(q_4q_5)^{{\bar{3}_c}}_{n_2^{2s_2+1}l_{2j_2}}\right)^{{{3}_c}}_{N_2^{2s_{3}+1}L_{2J_2}}\right)^{1_c}_{N_1^{2s_4+1}L_{1J_1}},
\eea
where {\cltb} denotes the color antitriplet state of the diquark, and $1_c$ represents the color singlet state of the pentaquark. [The superscript $1_c$ is often omitted, as the observed pentaquarks are colorless.]
The notation in Eq. (\ref{tetnot}) can also be written as $|n_1^{2s_1+1}l_{1j_1},n_2^{2s_2+1}l_{2j_2},N_2^{2s_{3}+1}L_{2J_2},N_1^{2s_4+1}L_{1J_1}\rangle$.
The diquark $(q_1q_2)$ is either $\{q_1q_2\}$ or $[q_1q_2]$, where $\{q_1q_2\}$ and $[q_1q_2]$ represent the permutation symmetric and antisymmetric flavor wave functions, respectively. $N_{1,2}=N_{r_{1,2}}+1$, where $N_{r_{1,2}}=0,\,1,\,\cdots$. $n_{1,2}=n_{r_{1,2}}+1$, where $n_{r_{1,2}}=0,\,1,\,\cdots$. $N_{r_1}$, $N_{r_2}$, $n_{r_1}$ and $n_{r_2}$ are the radial quantum numbers of the pentaquark, triquark, diquark 1, diquark 2, respectively.
$\vec{J}_1=\vec{L}_1+\vec{s}_4$, $\vec{s}_4=\vec{j}_1+\vec{J}_2$, $\vec{J}_2=\vec{s}_{3}+\vec{L}_2$, $\vec{s}_{3}=\vec{j}_2+\vec{s}_{\bar{q}_3}$, $\vec{j}_1=\vec{s}_1+\vec{l}_1$, $\vec{j}_2=\vec{s}_2+\vec{l}_2$.
$\vec{J}_1$, $\vec{J}_2$, $\vec{j}_1$ and $\vec{j}_2$ are the spins of pentaquark, triquark, diquark 1, diquark 2, respectively. $L_1$, $L_2$, $l_{1}$ and $l_2$ are the orbital quantum numbers of pentaquark, triquark, diquark 1 and diquark 2, respectively. $\vec{s}_{1}$ and $\vec{s}_{2}$ are the summed spin of quarks in the diquark $1$ and diquark $2$, respectively; $\vec{s}_{3}$ is the summed spin of antiquark $3$ and diquark $2$, and  $\vec{s}_{4}$ is the summed spin of diquark $1$ and triquark.

In the diquark-triquark picture, the fully heavy pentaquark $P_{cc\bar{c}bb}$ has three configurations: $(cc)(\bar{c}(bb))$, $(bb)(\bar{c}(cc))$, and $(bc)(\bar{c}(bc))$. In the first configuration, $P_{cc\bar{c}bb}$ consists of a diquark $(cc)$ and a triquark $(\bar{c}(bb))$, where the triquark is composed of a antiquark $\bar{c}$ and a diquark $(bb)$. In the second configuration, $P_{cc\bar{c}bb}$ is composed of a diquark $(bb)$ and a triquark $(\bar{c}(cc))$. In the third configuration, $P_{cc\bar{c}bb}$ consists two clusters: $(bc)$ and $(\bar{c}(bc))$. Due to mode mixing in the $(bc)(\bar{c}(bc))$ configuration, the corresponding {\rts} are complex. This configuration is not considered in the present work.

\subsection{{\rt} relations for the fully heavy pentaquarks}
The spinless Salpeter equation \cite{Godfrey:1985xj,Capstick:1986ter,Ferretti:2019zyh,
Bedolla:2019zwg,Durand:1981my,Durand:1983bg,Lichtenberg:1982jp,Jacobs:1986gv} is given by
\begin{eqnarray}\label{qsse}
M\Psi_{d,t,p}({\bfr})=H\Psi_{d,t,p}({\bfr}),\;\, H=\omega_1+\omega_2+V_{d,t,p},
\end{eqnarray}
where $M$ is the bound state mass (diquark, triquark or pentaquark). $\Psi_{d,t,p}({\bfr})$ denotes the wave function of the diquark, triquark, and pentaquark, respectively. $V_{d,t,p}$ represents the diquark, triquark, and pentaquark potentials, respectively (see Eq. (\ref{potv})). $\omega_1$ is the relativistic energy of constituent $1$ (quark, diquark, or antiquark), and $\omega_2$ is of constituent $2$ (quark, diquark, or triquark),
\bea\label{omega}
\omega_i=\sqrt{m_i^2+{\bf p}^2}=\sqrt{m_i^2-\Delta}\;\; (i=1,2).
\eea
Here, $m_1$ and $m_2$ are the effective masses of constituent $1$ and $2$, respectively.

The effect of the finite size of the diquark is treated differently. In Ref. \cite{Faustov:2021hjs}, the size of diquark is taken into account through appropriate form factors, while sometimes the diquark is treated as pointlike \cite{Ferretti:2019zyh,Lundhammar:2020xvw}.
For simplicity, the diquark is regarded as pointlike in the present work.

Following Refs. \cite{Ferretti:2019zyh,Bedolla:2019zwg,Ferretti:2011zz,Eichten:1974af}, we employ the potential
\begin{align}\label{potv}
V_{d,t,p}&=-\frac{3}{4}\left[V_c+{\sigma}r+C\right]
\left({\bf{F}_i}\cdot{\bf{F}_j}\right)_{d,t,p},
\end{align}
where $V_c\propto{1/r}$ is a color Coulomb potential or a Coulomb-like potential arising from one-gluon-exchange.
$C$ is a fundamental parameter \cite{Gromes:1981cb,Lucha:1991vn}. $\sigma$ is the string tension. The string tension $\sigma$ can be associated with the constituent quark masses \cite{Burns:2010qq,Lucha:1991vn} or treated as a common parameter for different quark masses \cite{Faustov:2021hjs,Faustov:2021qqf,Lucha:2000hg}. In this work, we adopt the latter approach. Although the masses for $J/\Psi$ and $\Psi'$ have undergone significant improvements since the 1970s, we use the value $\sigma=0.18$ ${\rm GeV}^2$ \cite{Godfrey:1985xj,Ebert:2007rn,Faustov:2021hjs}.
${\bf{F}_i}\cdot{\bf{F}_j}$ is the color-Casimir, with
\bea\label{mrcc}
\langle{(\bf{F}_i}\cdot{\bf{F}_j})_{d,t}\rangle=-\frac{2}{3},\quad
\langle{(\bf{F}_i}\cdot{\bf{F}_j})_{p}\rangle=-\frac{4}{3},
\eea
where $\langle{(\bf{F}_i}\cdot{\bf{F}_j})_{d,t,p}\rangle$ are the color-Casimirs for the diquark, the triquark, and the pentaquark, respectively.

The expression in brackets in Eq. (\ref{potv}) is the Cornell potential. The dynamics of hadrons are governed by a Hamiltonian that features one-gluon exchange dominance at short distances and a universal confinement potential independent of the hadron's quark content and dominating at large distances. This framework can accurately describe the spectroscopy of heavy mesons and baryons \cite{Ferretti:2019zyh,Godfrey:1985xj,Eichten:1974af}.
The Cornell form also persists in multiquark systems, including tetraquarks in the diquark-antidiquark picture \cite{Ferretti:2019zyh,Bedolla:2019zwg,Xie:2024dfe} and pentaquarks in the diquark-triquark picture \cite{Lebed:2015tna}.
In the diquark picture, a baryon consists of one quark in $3_c$ and a diquark in {\cltb}, while a pentaquark is composed of a diquark in {\cltb} and a triquark in $3_c$. The Cornell approach reduces the pentaquark modeling (via diquark-triquark interaction) problem to a meson-like system, where a meson is composed of a quark in $3_c$ and an antiquark in {\cltb} \cite{Ferretti:2019zyh,Bedolla:2019zwg,Lebed:2015tna,Brodsky:2014xia,Song:2024bkj}.
The color-Casimir values corresponding to quark-antiquark interactions for mesons, quark-diquark interactions for baryons, antidiquark-diquark interactions for tetraquarks, and triquark-diquark interactions for pentaquarks are all $-4/3$.

For heavy-heavy systems, where $m_{1},m_2{\gg}{\bfpa}$, Eq. (\ref{qsse}) reduces to
\bea\label{qssenrr}
M\Psi_{d,t,p}({\bfr})=\left[(m_1+m_2)+\frac{{\bfp}^2}{2\mu}+V_{d,t,p}\right]\Psi_{d,t,p}({\bfr}),
\eea
where
\bea\label{rdmu}
\mu=m_1m_2/(m_1+m_2).
\eea
By employing the Bohr-Sommerfeld quantization approach \cite{brau:04bs} and using Eqs. (\ref{potv}) and (\ref{qssenrr}), we obtain the following parametrized relation  \cite{Chen:2022flh,Chen:2021kfw}
\begin{align}\label{massform}
M=&m_R+\beta_x(x+c_{0x})^{2/3},\nonumber\\
x=&l_1,l_2,\,n_{r_1},\,n_{r_2},\,L_1,\,L_2,\,N_{r_1}\,N_{r_2},
\end{align}
with
\bea\label{parabm}
\beta_x=c_{fx}c_xc_c,\quad m_R=m_1+m_2+C',
\eea
where
\bea\label{cprime}
C'=\left\{\begin{array}{cc}
C/2, & \text{diquarks,\,triquarks}, \\
C, & \text{pentaquarks}.
\end{array}\right.
\eea
\bea\label{sigma}
\sigma'=\left\{\begin{array}{cc}
\sigma/2, & \text{diquarks,\,triquarks}, \\
\sigma, & \text{pentaquarks}.
\end{array}\right.
\eea
$c_{x}$ and $c_c$ are
\begin{align}\label{cxcons}
c_c=\left(\frac{\sigma'^2}{\mu}\right)^{1/3},&\quad c_{l_1,l_2,L_1,L_2}=\frac{3}{2},\nonumber\\
c_{n_{r_1},n_{r_2},N_{r_1},N_{r_2}}&=\frac{\left(3\pi\right)^{2/3}}{2}.
\end{align}
Theoretically, $c_{fx}$ equals one but int practice is fitted to data.
In Eq. (\ref{massform}), $m_1$, $m_2$, $c_x$ and $\sigma$ are universal parameters for heavy-heavy systems. $c_{0x}$ varies with different {\rts}.

Using Eqs. (\ref{massform}), (\ref{parabm}), (\ref{cprime}), (\ref{sigma}), and (\ref{cxcons}), we obtain the {\rt} relations for fully heavy pentaquarks:
\begin{align}\label{ppt2q}
M&=m_{R_{\lambda_1}}+\beta_{x_{\lambda_1}}(x_{\lambda_1}+c_{0x_{\lambda_1}})^{2/3}\;(x_{\lambda_1}=L_1,\,N_{r_1}),\nonumber\\
m_{R_{\lambda_1}}&=M_{d_1}+M_{t}+C,\nonumber\\
M_{d_1}&=m_{R_{\rho_1}}+\beta_{x_{\rho_1}}(x_{\rho_1}+c_{0x_{\rho_1}})^{2/3}\;(x_{\rho_1}=l_1,\,n_{r_1}),\nonumber\\
m_{R_{\rho_1}}&=m_{q_1}+m_{q_2}+C/2,\nonumber\\
M_{t}&=m_{R_{\lambda_2}}+\beta_{x_{\lambda_2}}(x_{\lambda_2}+c_{0x_{\lambda_2}})^{2/3}\;(x_{\lambda_2}=L_2,\,N_{r_2}),\nonumber\\
m_{R_{\lambda_2}}&=M_{d_2}+m_{q_3}+C/2,\nonumber\\
M_{d_2}&=m_{R_{\rho_2}}+\beta_{x_{\rho_2}}(x_{\rho_2}+c_{0x_{\rho_2}})^{2/3}\;(x_{\rho_2}=l_2,\,n_{r_2}),\nonumber\\
m_{R_{\rho_2}}&=m_{q_4}+m_{q_5}+C/2,
\end{align}
where
\begin{align}\label{pppa2qQ}
\mu_{\lambda_1}&=\frac{M_{d_1}M_{t}}{M_{d_1}+M_{t}},\;
\mu_{\rho_1}=\frac{m_{q_1}m_{q_2}}{m_{q_1}+m_{q_2}},\nonumber\\
\beta_{L_1}&=\frac{3}{2}\left(\frac{\sigma^2}{\mu_{\lambda_1}}\right)^{1/3}c_{fL_1},\; \beta_{N_{r_1}}=\frac{(3\pi)^{2/3}}{2}\left(\frac{\sigma^2}{\mu_{\lambda_1}}\right)^{1/3}c_{fN_{r_1}},\nonumber\\
\beta_{l_1}&=\frac{3}{2}\left(\frac{\sigma^2}{4\mu_{\rho_1}}\right)^{1/3}c_{fl_1},\; \beta_{n_{r_1}}=\frac{(3\pi)^{2/3}}{2}\left(\frac{\sigma^2}{4\mu_{\rho_1}}\right)^{1/3}c_{fn_{r_1}},\nonumber\\
\mu_{\lambda_2}&=\frac{M_{d_2}m_{q_3}}{M_{d_2}+m_{q_3}},\;
\mu_{\rho_2}=\frac{m_{q_4}m_{q_5}}{m_{q_4}+m_{q_5}},\nonumber\\
\beta_{L_2}&=\frac{3}{2}\left(\frac{\sigma^2}{4\mu_{\lambda_2}}\right)^{1/3}c_{fL_2},\; \beta_{N_{r_2}}=\frac{(3\pi)^{2/3}}{2}\left(\frac{\sigma^2}{4\mu_{\lambda_2}}\right)^{1/3}c_{fN_{r_2}},\nonumber\\
\beta_{l_2}&=\frac{3}{2}\left(\frac{\sigma^2}{4\mu_{\rho_2}}\right)^{1/3}c_{fl_2},\; \beta_{n_{r_2}}=\frac{(3\pi)^{2/3}}{2}\left(\frac{\sigma^2}{4\mu_{\rho_2}}\right)^{1/3}c_{fn_{r_2}}.
\end{align}
Here, $M$, $M_t$, $M_{d_1}$, $M_{d_2}$, $m_{q_1}$, $m_{q_2}$, $m_{q_3}$, $m_{q_4}$, and $m_{q_5}$ are the masses of the pentaquark, triquark, diquark $1$, diquark $2$, quark $1$, quark $2$, quark $3$, quark $4$, and quark $5$, respectively. 

In Eq. (\ref{ppt2q}), $M_{d_1}$ and $M_{d_2}$ give the diquark {\rts} \cite{Feng:2023txx}, $M_{d_1}$, $M_{d_2}$ and $M_t$ can be used to construct the triquark {\rts} \cite{Song:2024bkj}. $M_{d_1}$, $M_{d_2}$, $M_t$ and $M$ generate the {\rts} for the fully heavy pentaquarks.
According to Eqs. (\ref{ppt2q}) and (\ref{pppa2qQ}), we have
\bea
M=M_{d_1}+M_{t}+C+\beta_{x_{\lambda_1}}(x_{\lambda_1}+c_{0x_{\lambda_1}})^{2/3}\;(x_{\lambda_1}=L_1,\,N_{r_1})
\eea
when diquark $1$ and the triquark are regarded as structureless constituents, the binding energies of the pentaquark is $\epsilon=C+\beta_{x_{\lambda_1}}(x_{\lambda_1}+c_{0x_{\lambda_1}})^{2/3}$. When the triquark is taken as a bound state composed a diquark and an antiquark, and diquarks $1$ and $2$ are each regarded as bound states of two heavy quarks, we have
\begin{align}\label{rtpf}
M=&m_{q_1}+m_{q_2}+m_{q_3}+m_{q_4}+m_{q_5}+\frac{5}{2}C \nonumber\\
&+\beta_{x_{\lambda_1}}(x_{\lambda_1}+c_{0x_{\lambda_1}})^{2/3}
+\beta_{x_{\rho_1}}(x_{\rho_1}+c_{0x_{\rho_1}})^{2/3}\nonumber\\
&+\beta_{x_{\lambda_2}}(x_{\lambda_2}+c_{0x_{\lambda_2}})^{2/3}
+\beta_{x_{\rho_2}}(x_{\rho_2}+c_{0x_{\rho_2}})^{2/3}.
\end{align}
From the pentaquark {\rt} relations [Eq. (\ref{rtpf})], it can be seen that there are four series of masses and, correspondingly, four series of {\rts} for fully heavy pentaquarks: the $\lambda_1$-trajectories (with $x_{\lambda_2}$, $x_{\rho_1}$, and $x_{\rho_2}$ fixed); the $\lambda_2$-trajectories (with $x_{\lambda_1}$, $x_{\rho_1}$, and $x_{\rho_2}$ fixed); the $\rho_1$-trajectories (with $x_{\lambda_1}$, $x_{\lambda_2}$, and $x_{\rho_2}$ fixed); and the $\rho_2$-trajectories (with $x_{\lambda_1}$, $x_{\lambda_2}$, and $x_{\rho_1}$ fixed).

For later convenience, we refer to the {\rts} calculated from Eqs. (\ref{ppt2q}) and (\ref{pppa2qQ}), or from Eqs. (\ref{rtpf}) and (\ref{pppa2qQ}), as the complete forms of the {\rts}. The obtained constant and the mode under consideration are referred to the main part of the {\rts}. For example, when considering the $\rho_1$-trajectories, $\beta_{x_{\lambda_2}}(x_{\lambda_2}+c_{0x_{\lambda_2}})^{2/3}$ and $\beta_{x_{\rho_2}}(x_{\rho_2}+c_{0x_{\rho_2}})^{2/3}$ are constants, while $\beta_{x_{\lambda_1}}(x_{\lambda_1}+c_{0x_{\lambda_1}})^{2/3}$ becomes a function of $x_{\rho_1}$ (through the dependence in $\beta_{x_{\lambda_1}}$). Therefore, the main part of the $\rho_1$-trajectories is $\widetilde{m}_R+\beta_{x_{\rho_1}}(x_{\rho_1}+c_{0x_{\rho_1}})^{2/3}$, where $\widetilde{m}_R=m_{q_1}+m_{q_2}+m_{q_3}+m_{q_4}+m_{q_5}+\frac{5}{2}C +\beta_{x_{\lambda_2}}(x_{\lambda_2}+c_{0x_{\lambda_2}})^{2/3}
+\beta_{x_{\rho_2}}(x_{\rho_2}+c_{0x_{\rho_2}})^{2/3}$. The difference between the complete form of the $\rho_1$-{\tr} and its main part is $\beta_{x_{\lambda_1}}(x_{\lambda_1}+c_{0x_{\lambda_1}})^{2/3}$.

In the diquark-triquark picture, the problem of modeling pentaquarks (via diquark-triquark interaction) reduces to a meson-like system \cite{Lebed:2015tna,Brodsky:2014xia}. In the meson case, spin-dependent interactions are relativistic corrections and are small compared to the Cornell potential. These contributions do not alter the behavior of meson Regge trajectories. Similarly, spin-dependent interactions between the diquark and triquark do not affect the behavior of pentaquark Regge trajectories. In this work, we focus on the Regge trajectories of fully heavy pentaquarks and do not consider spin-dependent effects, which will be addressed in future studies.

\section{{\rts} for the pentaquark $P_{cc\bar{c}bb}$}\label{sec:rts}

\subsection{Parameters}

\begin{table}[!phtb]
\caption{The values of parameters \cite{Feng:2023txx,Faustov:2021qqf}.}  \label{tab:parmv}
\centering
\begin{tabular*}{0.45\textwidth}{@{\extracolsep{\fill}}cc@{}}
\hline\hline
          & $m_{c}=1.55\; {\gev}$, \; $m_b=4.88\; {\gev}$, \\
          & $\sigma=0.18\; {\gev^2}$,\; $C=-0.3\; {\gev}$, \\
          & $c_{fn_{r}}=1.0$,\; $c_{fl}=1.17$        \\
$(cc)$    & $c_{0n_{r}}(1^3s_1)=0.205$,\quad $c_{0{l}}(1^3s_1)=0.337$,\\
$(bb)$    & $c_{0n_{r}}(1^3s_1)=0.01$,\quad  $c_{0{l}}(1^3s_1)=0.001$.\\
\hline
\hline
\end{tabular*}
\end{table}

The parameter values are listed in Table \ref{tab:parmv}. The values of $m_b$, $m_c$, $\sigma$ and $C$ are adopted directly from Ref. \cite{Faustov:2021qqf}. $c_{fx}$ and $c_{0x}$ for the $\rho$-mode are extracted by fitting the {\rts} for doubly heavy mesons, and are subsequently used as input to fit the {\rts} for doubly heavy diquarks. The parameter $c_{fx}$ is universal for all doubly heavy diquark {\rts}, whereas $c_{0x}$ varies with different diquark {\rts} \cite{Feng:2023txx}.
The parameters for the $\lambda$-mode are determined by the following relations \cite{Xie:2024dfe}
\begin{eqnarray}
c_{fL_1,fL_2}=&1.116 + 0.013\mu_{\lambda_1,\lambda_2},\nonumber\\
 c_{0L_1,0L_2}=&0.540- 0.141\mu_{\lambda_1,\lambda_2}, \nonumber\\
c_{fN_{r_1},fN_{r_2}}=&1.008 + 0.008\mu_{\lambda_1,\lambda_2},\nonumber\\ c_{0N_{r_1},0N_{r_2}}=&0.334 - 0.087\mu_{\lambda_1,\lambda_2},\label{fitcfxnr}
\end{eqnarray}
where $\mu_{\lambda_1}$ and $\mu_{\lambda_2}$ are the reduced masses (see Eq. (\ref{pppa2qQ})). The relations in Eq. (\ref{fitcfxnr}) are obtained by fitting the mesons, baryons, and tetraquarks. These relations are employed as a provisional method until a better approach becomes available. Validation can be achieved by comparing the fitted results for pentaquarks with theoretical values obtained from other approaches and available experimental data.

\subsection{$\rho$-{\trs} for the pentaquark $P_{cc\bar{c}bb}$}\label{subsec:rts}

\begin{table*}[!phtb]
\caption{The masses of the radially and orbitally $\rho_1$- and $\rho_2$-excited states of the pentaquark $P_{cc\bar{c}bb}$. Both configurations $(cc)(\bar{c}(bb))$ and $(bb)(\bar{c}(cc))$ are considered. The notation in Eq. (\ref{tetnot}) is rewritten as $|n_1^{2s_1+1}l_{1j_1},n_2^{2s_2+1}l_{2j_2},N_2^{2s_{3}+1}L_{2J_2},N_1^{2s_4+1}L_{1J_1}\rangle$. $|n_1^{2s_1+1}l_{1j_1},n_2^{2s_2+1}l_{2j_2},N_2L_{2},N_1L_{1}\rangle$ denotes the spin-averaged states. Eqs. (\ref{ppt2q}), (\ref{pppa2qQ}) and (\ref{fitcfxnr}) are used in the calculation. States marked with $(\times)$ do not exist.}  \label{tab:rho}
\centering
\begin{tabular*}{1.0\textwidth}{@{\extracolsep{\fill}}ccc@{}}
\hline\hline
  $|n_1^{2s_1+1}l_{1j_1},n_2^{2s_2+1}l_{2j_2},N_2L_{2},N_1L_{1}\rangle$        & $(cc)(\bar{c}(bb))$ [{\gev}] &  $(bb)(\bar{c}(cc))$ [{\gev}]  \\
\hline
 $|1^3s_1, 1^3s_1, 1S, 1S\rangle$  & 14.13   & 14.10    \\
 $|2^3s_1, 1^3s_1, 1S, 1S\rangle$  & 14.49   & 14.41  \\
 $|3^3s_1, 1^3s_1, 1S, 1S\rangle$  & 14.75   & 14.61   \\
 $|4^3s_1, 1^3s_1, 1S, 1S\rangle$  & 14.98   & 14.77   \\
 $|5^3s_1, 1^3s_1, 1S, 1S\rangle$  & 15.18   & 14.91    \\
 $|1^3s_1, 1^3s_1, 1S, 1S\rangle$  & 14.13   & 14.10 \\
 $|1^3s_1, 2^3s_1, 1S, 1S\rangle$  & 14.45   & 14.46  \\
 $|1^3s_1, 3^3s_1, 1S, 1S\rangle$  & 14.64   & 14.72  \\
 $|1^3s_1, 4^3s_1, 1S, 1S\rangle$  & 14.80   & 14.95 \\
 $|1^3s_1, 5^3s_1, 1S, 1S\rangle$  & 14.95   & 15.15  \\
 $|1^3s_1, 1^3s_1, 1S, 1S\rangle$  & 14.14   & 14.08    \\
 $|1^3p_2, 1^3s_1, 1S, 1S\rangle(\times)$  & 14.41   & 14.34  \\
 $|1^3d_3, 1^3s_1, 1S, 1S\rangle$  & 14.61   & 14.49   \\
 $|1^3f_4, 1^3s_1, 1S, 1S\rangle(\times)$  & 14.78   & 14.62   \\
 $|1^3g_5, 1^3s_1, 1S, 1S\rangle$  & 14.94   & 14.74    \\
 $|1^3s_1, 1^3s_1, 1S, 1S\rangle$  & 14.12   & 14.11 \\
 $|1^3s_1, 1^3p_2, 1S, 1S\rangle(\times)$  & 14.37   & 14.38  \\
 $|1^3s_1, 1^3d_3, 1S, 1S\rangle$  & 14.53   & 14.58  \\
 $|1^3s_1, 1^3f_4, 1S, 1S\rangle(\times)$  & 14.66   & 14.75 \\
 $|1^3s_1, 1^3g_5, 1S, 1S\rangle$  & 14.77   & 14.91  \\
\hline\hline
\end{tabular*}
\end{table*}

\begin{figure*}[!phtb]
\centering
\subfigure[]{\label{subfigure:cfa}\includegraphics[scale=0.45]{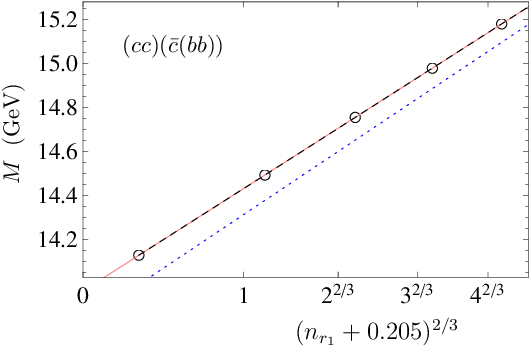}}
\subfigure[]{\label{subfigure:cfa}\includegraphics[scale=0.45]{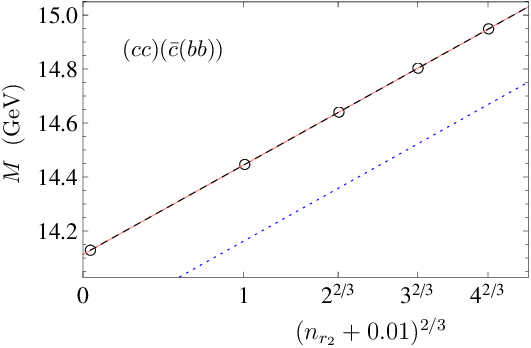}}
\subfigure[]{\label{subfigure:cfa}\includegraphics[scale=0.45]{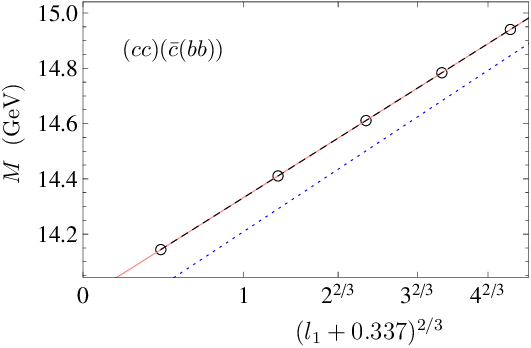}}
\subfigure[]{\label{subfigure:cfa}\includegraphics[scale=0.45]{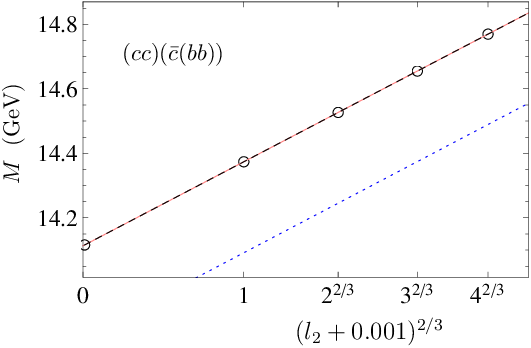}}
\subfigure[]{\label{subfigure:cfa}\includegraphics[scale=0.45]{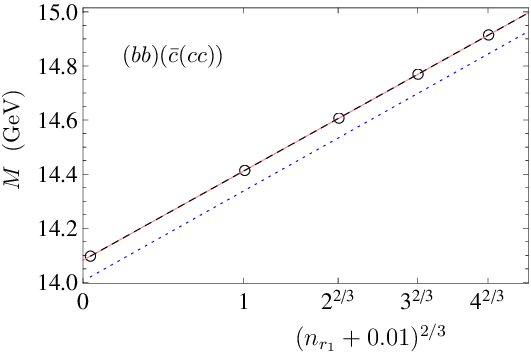}}
\subfigure[]{\label{subfigure:cfa}\includegraphics[scale=0.45]{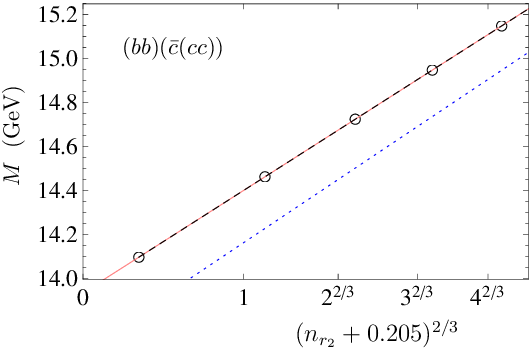}}
\subfigure[]{\label{subfigure:cfa}\includegraphics[scale=0.45]{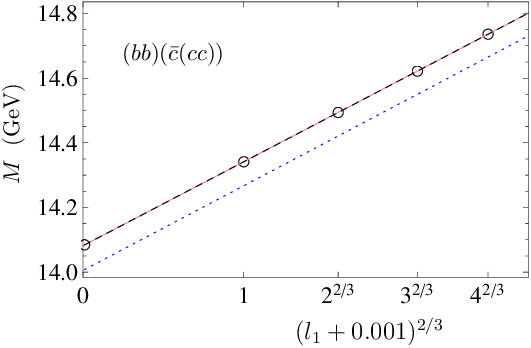}}
\subfigure[]{\label{subfigure:cfa}\includegraphics[scale=0.45]{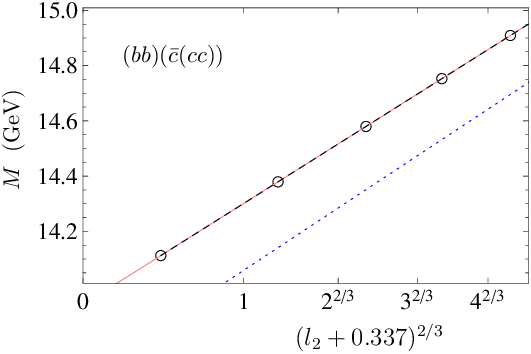}}
\caption{The $\rho_1$- and $\rho_2$-{\trs} for the pentaquark $P_{cc\bar{c}bb}$ in the configurations $(cc)(\bar{c}(bb))$ and $(bb)(\bar{c}(cc))$. $n_{r_1}$ and ${l_1}$ are the radial and orbital quantum numbers for the $\rho_1$-mode, respectively, while $n_{r_2}$ and ${l_2}$ are the corresponding numbers for the $\rho_2$-mode.  Circles represent the predicted data listed in Table \ref{tab:rho}. The black dashed lines correspond to the $\rho$-{\trs} for the complete forms, obtained from Eqs. (\ref{ppt2q}) and (\ref{pppa2qQ}) or from Eqs. (\ref{rtpf}) and (\ref{pppa2qQ}). The pink lines correspond to the fitted formulas, obtained by linearly fitting the predicted data in Table \ref{tab:rho}; these formulas are listed in Table \ref{tab:formulas}. The blue dotted lines correspond to the main parts of the complete form, which are also listed in Table \ref{tab:formulas}.}\label{fig:rho}
\end{figure*}

In this subsection, the pentaquark $P_{cc\bar{c}bb}$ in the configurations $(cc)(\bar{c}(bb))$ and $(bb)(\bar{c}(cc))$ is considered. By using Eqs. (\ref{ppt2q}), (\ref{pppa2qQ}) and (\ref{fitcfxnr}) and parameters in Table \ref{tab:parmv}, the spin-averaged masses of the radially and orbitally $\rho_1$- and $\rho_2$-excited states are calculated. The calculated results are listed in Table \ref{tab:rho}. More results can be easily obtained in a similar manner.
When calculating the $\rho_1$-modes, all other modes are kept in their ground states, and the parameters used correspond to the radial ground states. By a similar procedure, the spin-averaged masses of the $\rho_2$-excited states can be determined.

In the configuration $(cc)(\bar{c}(bb))$, the masses of the radially and orbitally $\rho_1$-excited states are greater than those of the corresponding $\rho_2$-excited states (see Table \ref{tab:rho}). Conversely, in the configuration $(bb)(\bar{c}(cc))$, the masses of the $\rho_2$-excited states exceed those of the $\rho_1$-excited states. This is because the masse increase from excitation of the diquark $(cc)$ is greater than that from excitation of the diquark $(bb)$.
Accordingly, the following inequalities hold for the pentaquark $P_{cc\bar{c}bb}$:
\begin{align}
M\left(\rho_1, (cc)(\bar{c}(bb))\right)>
M\left(\rho_2, (cc)(\bar{c}(bb))\right),\nonumber\\
M\left(\rho_2, (bb)(\bar{c}(cc))\right)>
M\left(\rho_1, (bb)(\bar{c}(cc))\right),\nonumber\\
M\left(\rho_1, (cc)(\bar{c}(bb))\right)>
M\left(\rho_1, (bb)(\bar{c}(cc))\right),\nonumber 
\end{align}
\begin{align}\label{masseq}
M\left(\rho_2, (bb)(\bar{c}(cc))\right)>
M\left(\rho_2, (cc)(\bar{c}(bb))\right),
\end{align}
where $M\left(\rho_1, (bb)(\bar{c}(cc))\right)$ denotes the masses of the radially or orbitally $\rho_1$-mode excited state of $P_{cc\bar{c}bb}$ in the configuration $(bb)(\bar{c}(cc))$; the other notations in Eq. (\ref{masseq}) are similarly defined.
The masses of the $\rho_1$-excited ($\rho_2$-excited) states in the configuration $(cc)(\bar{c}(bb))$ are close to those of the $\rho_2$-excited ($\rho_1$-excited) states in the configuration $(bb)(\bar{c}(cc))$, i.e., $M\big(\rho_1, (cc)(\bar{c}(bb))\big){\approx}M\big(\rho_2, (bb)(\bar{c}(cc))\big)$, $M\big(\rho_1, (bb)(\bar{c}(cc))\big){\approx}M\big(\rho_2, (cc)(\bar{c}(bb))\big)$. Possible mixings between these states are neglected here.

Unlike the {\rts} for light mesons, which have a simple linear form, the {\rts} for fully heavy pentaquarks become tedious when plotted in the $(M^2,x)$ plane. For example, the fitted $\rho_1$-{\rt} changes from Eq. (\ref{fra}) to
\begin{align}\label{msqr}
M^2=&195.052+0.216964(0.205 + n_{r_1})^{4/3}\nonumber\\
&+13.0107(0.205 + n_{r_1})^{2/3}.
\end{align}
The squared complete form of $\rho_1$-{\rt} [see Eq. (\ref{fulla})] will be even more complicated.
Furthermore, the behavior of the Regge trajectory in terms of $M^2$ is less obvious and transparent than in terms of $M$. For the fitted $\rho_1$-{\rt}, $M{\sim}n_{r_1}^{2/3}$, while $M^2{\sim}n_{r_1}^{\nu}$ with $\nu$ not being clearly defined but greater than $2/3$, and $\nu$ varies for different {\trs}. When Eq. (\ref{msqr}) is approximated as $M^2=m'_R+\beta'(x+c_0)^{2/3}$ because the term $(0.205 + n_{r_1})^{2/3}$ plays the dominant role, the resulting estimates  become less accurate compared with those obtained using $M=m_R+\beta(x+c_0)^{2/3}$.

Inspired by Ref. \cite{Burns:2010qq}, we employ the $(M,\,(x+c_0)^{2/3})$ plane, rather than the $(M^2,\,x)$ plane, to plot the $\rho_1$- and $\rho_2$-{\trs} for the pentaquark $P_{cc\bar{c}bb}$ in the configurations $(cc)(\bar{c}(bb))$ and $(bb)(\bar{c}(cc))$ (see Fig. \ref{fig:rho}). The full forms of the $\rho_1$- and $\rho_2$-{\trs}, directly calculated from Eqs. (\ref{ppt2q}) and (\ref{pppa2qQ}) or from Eqs. (\ref{rtpf}) and (\ref{pppa2qQ}), are indicated by the black dashed lines in Fig. \ref{fig:rho}. For illustration, the full expressions are presented in Eqs. (\ref{fulla}) and (\ref{fullb}). These expressions are generally lengthy and cumbersome, and their complexity varies with different {\rts}.
By performing a linear fit to the calculated data in Table \ref{tab:rho}, we obtain the fitted formulas listed in Table \ref{tab:formulas}. In Fig. \ref{fig:rho}, the fitted formulas are represented by the pink lines.
As shown in Fig. \ref{fig:rho}, these fitted formulas (pink lines) nearly overlap with the complete forms of the $\rho_1$- and $\rho_2$-{\trs} (black dashed lines). This demonstrates that the complex full forms of the $\rho$-{\rts} can be well approximated by the simple fitted formulas. Accordingly, the complete forms of the $\rho_1$- and $\rho_2$-{\trs} both exhibit the behavior $M{\sim}x_{\rho_1}^{2/3},\;x_{\rho_2}^{2/3}$, where $x_{\rho_1}=n_{r_1},\;l_1$ and $x_{\rho_2}=n_{r_2},\;l_2$. The main parts of the full forms of the $\rho_1$- and $\rho_2$-{\trs} (listed in Table \ref{tab:formulas} and indicated by the blue dotted lines in Fig. \ref{fig:rho}) show significant deviations from the full forms (black dashed lines). For both the radial and orbital Regge trajectories, and for both configurations $(cc)(\bar{c}(bb))$ and $(bb)(\bar{c}(cc))$, the main parts of the $\rho_2$-{\trs} deviate more from the complete forms than do those of the $\rho_1$-{\trs}.
In conclusion, the $\rho$-{\trs} are presented in Fig. \ref{fig:rho}, and the Chew-Frautschi plots clearly demonstrate the behavior of {\trs} (see the footnote on the first page). The tedious complete forms of the $\rho$-{\trs}--in which the {\rt} behavior is not obvious--can be well approximated by the simple fitted forms, in which the {\rt} behavior is apparent. Furthermore, the differences between the main parts and the complete forms vary among different series of Regge trajectories.

\subsection{$\lambda$-{\trs} for the pentaquark $P_{cc\bar{c}bb}$}\label{subsec:rts}

\begin{table*}[!phtb]
\caption{Same as Table \ref{tab:rho} except for the $\lambda_1$- and $\lambda_2$-excited states.}  \label{tab:lambda}
\centering
\begin{tabular*}{1.0\textwidth}{@{\extracolsep{\fill}}ccc@{}}
\hline\hline
  $|n_1^{2s_1+1}l_{1j_1},n_2^{2s_2+1}l_{2j_2},N_2L_{2},N_1L_{1}\rangle$        & $(cc)(\bar{c}(bb))$ [{\gev}]  &  $(bb)(\bar{c}(cc))$ [{\gev}]  \\
\hline
 $|1^3s_1, 1^3s_1, 1S, 1S\rangle$  & 14.13   & 14.10    \\
 $|1^3s_1, 1^3s_1, 2S, 1S\rangle$  & 14.45   & 14.43  \\
 $|1^3s_1, 1^3s_1, 3S, 1S\rangle$  & 14.68   & 14.67   \\
 $|1^3s_1, 1^3s_1, 4S, 1S\rangle$  & 14.88   & 14.88   \\
 $|1^3s_1, 1^3s_1, 5S, 1S\rangle$  & 15.06   & 15.06    \\
 $|1^3s_1, 1^3s_1, 1S, 1S\rangle$  & 14.13   & 14.10    \\
 $|1^3s_1, 1^3s_1, 1S, 2S\rangle$  & 14.58   & 14.54  \\
 $|1^3s_1, 1^3s_1, 1S, 3S\rangle$  & 14.89   & 14.83   \\
 $|1^3s_1, 1^3s_1, 1S, 4S\rangle$  & 15.15   & 15.08   \\
 $|1^3s_1, 1^3s_1, 1S, 5S\rangle$  & 15.39   & 15.30    \\
 $|1^3s_1, 1^3s_1, 1S, 1S\rangle$  & 14.13   & 14.10    \\
 $|1^3s_1, 1^3s_1, 1P, 1S\rangle$  & 14.36   & 14.33  \\
 $|1^3s_1, 1^3s_1, 1D, 1S\rangle$  & 14.53   & 14.51   \\
 $|1^3s_1, 1^3s_1, 1F, 1S\rangle$  & 14.67   & 14.66   \\
 $|1^3s_1, 1^3s_1, 1G, 1S\rangle$  & 14.80   & 14.80    \\
 $|1^3s_1, 1^3s_1, 1S, 1S\rangle$  & 14.13   & 14.10    \\
 $|1^3s_1, 1^3s_1, 1S, 1P\rangle$  & 14.45   & 14.42  \\
 $|1^3s_1, 1^3s_1, 1S, 1D\rangle$  & 14.68   & 14.64   \\
 $|1^3s_1, 1^3s_1, 1S, 1F\rangle$  & 14.88   & 14.82   \\
 $|1^3s_1, 1^3s_1, 1S, 1G\rangle$  & 15.06   & 14.99    \\
\hline\hline
\end{tabular*}
\end{table*}

\begin{figure*}[!phtb]
\centering
\subfigure[]{\label{subfigure:cfa}\includegraphics[scale=0.45]{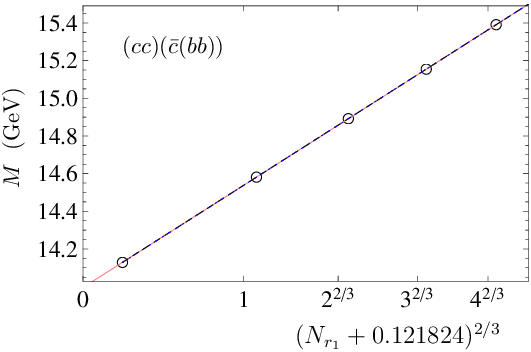}}
\subfigure[]{\label{subfigure:cfa}\includegraphics[scale=0.45]{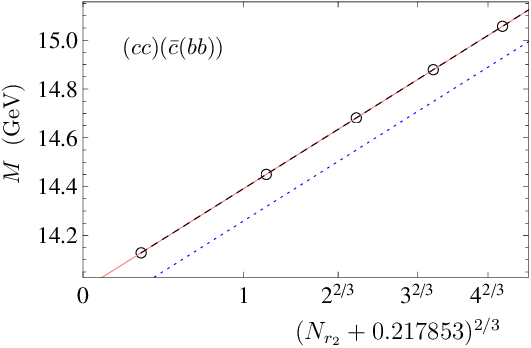}}
\subfigure[]{\label{subfigure:cfa}\includegraphics[scale=0.45]{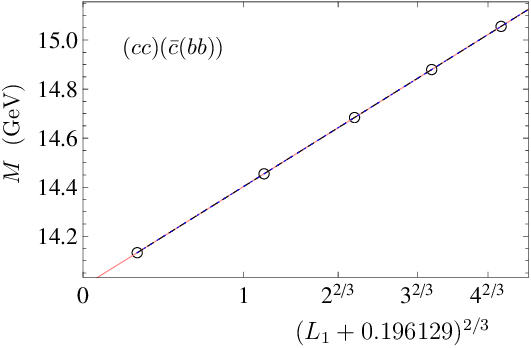}}
\subfigure[]{\label{subfigure:cfa}\includegraphics[scale=0.45]{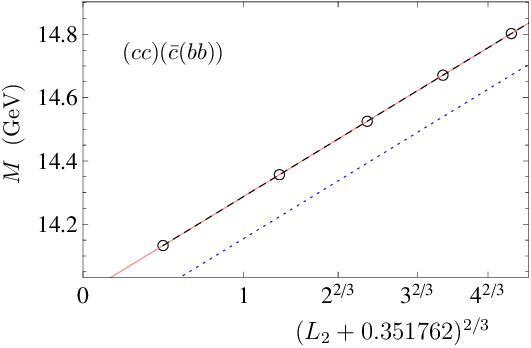}}
\subfigure[]{\label{subfigure:cfa}\includegraphics[scale=0.45]{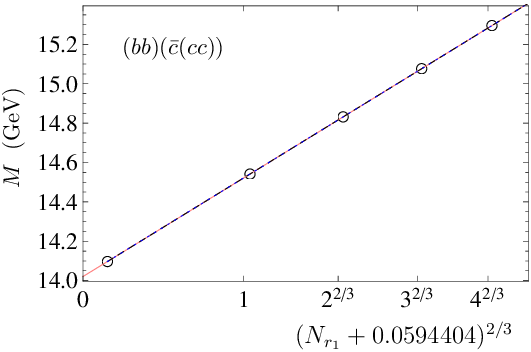}}
\subfigure[]{\label{subfigure:cfa}\includegraphics[scale=0.45]{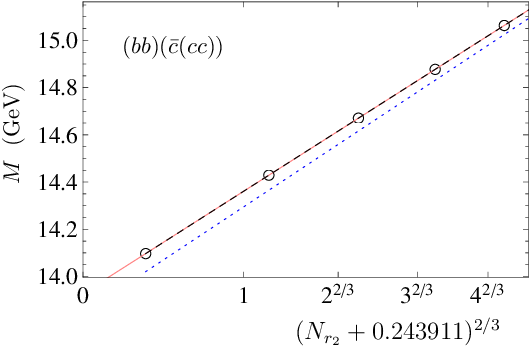}}
\subfigure[]{\label{subfigure:cfa}\includegraphics[scale=0.45]{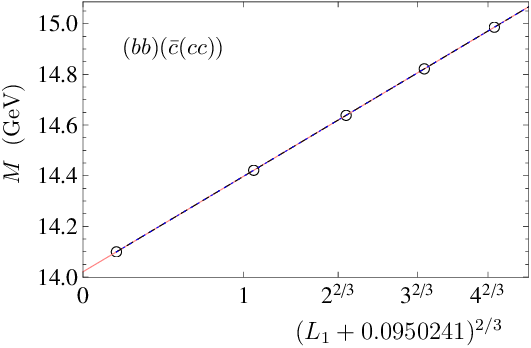}}
\subfigure[]{\label{subfigure:cfa}\includegraphics[scale=0.45]{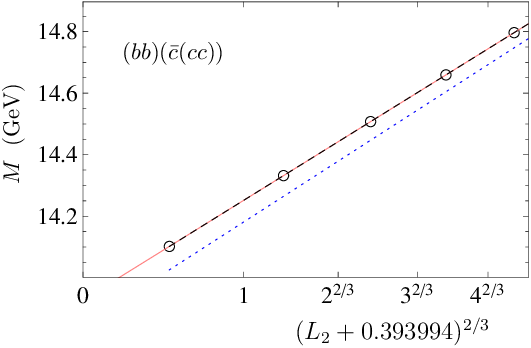}}
\caption{The $\lambda_1$- and $\lambda_2$-{\trs} for the pentaquark $P_{cc\bar{c}bb}$ in configurations $(cc)(\bar{c}(bb))$ and $(bb)(\bar{c}(cc))$. $N_{r_1}$ and ${L_1}$ are the radial and orbital quantum numbers for the $\lambda_1$-mode, respectively. $N_{r_2}$ and ${L_2}$ are the radial and orbital quantum numbers for the $\lambda_2$-mode, respectively.  Circles represent the predicted data listed in Table \ref{tab:lambda}. The black dashed lines correspond to the $\lambda$-{\trs} for the complete forms, obtained from Eqs. (\ref{ppt2q}) and (\ref{pppa2qQ}) or from Eqs. (\ref{rtpf}) and (\ref{pppa2qQ}). The pink lines correspond to the fitted formulas, obtained by linearly fitting the calculated data in Table \ref{tab:lambda}; these formulas are listed in Table \ref{tab:formulas}. The blue dotted lines are for the main parts of the full forms, which are also listed in Table \ref{tab:formulas}.}\label{fig:lambda}
\end{figure*}

In this subsection, the $\lambda_1$-mode and $\lambda_2$-mode of the pentaquark $P_{cc\bar{c}bb}$ in the configurations $(cc)(\bar{c}(bb))$ and $(bb)(\bar{c}(cc))$ are considered. Using Eqs. (\ref{ppt2q}), (\ref{pppa2qQ}) and (\ref{fitcfxnr}), together with the parameters in Table \ref{tab:parmv}, the masses of the radially and orbitally $\lambda_1$- and $\lambda_2$-excited states are calculated. The calculated results are listed in Table \ref{tab:lambda}.
When calculating the $\lambda_1$  modes, all other modes are kept in their ground states, and the parameters used correspond to the radial ground states of those modes. By a similar procedure, the masses of the $\lambda_2$-excited states can be determined.

In both configurations $(cc)(\bar{c}(bb))$ and $(bb)(\bar{c}(cc))$, the masses of the radially and orbitally $\lambda_1$-excited states are greater than those of the corresponding $\lambda_2$-excited states (see Table \ref{tab:lambda}).
Accordingly, the following relations hold for the pentaquark $P_{cc\bar{c}bb}$
\begin{align}\label{masseql}
M\left(\lambda_1, (cc)(\bar{c}(bb))\right)>
M\left(\lambda_2, (cc)(\bar{c}(bb))\right),\nonumber\\
M\left(\lambda_1, (cc)(\bar{c}(bb))\right)>
M\left(\lambda_2, (bb)(\bar{c}(cc))\right),\nonumber\\
M\left(\lambda_1, (bb)(\bar{c}(cc))\right)>
M\left(\lambda_2, (bb)(\bar{c}(cc))\right),\nonumber\\
M\left(\lambda_1, (bb)(\bar{c}(cc))\right)>
M\left(\lambda_2, (cc)(\bar{c}(bb))\right),
\end{align}
where $M\left(\lambda_1, (bb)(\bar{c}(cc))\right)$ denotes the masses of the radially or orbitally $\lambda_1$-excited state of $P_{cc\bar{c}bb}$ in configuration $(bb)(\bar{c}(cc))$; the other notations in Eq. (\ref{masseql}) are similarly defined.
The masses of the $\lambda_1$-excited ($\lambda_2$-excited) states in the configuration $(cc)(\bar{c}(bb))$ are close to those of the $\lambda_1$-excited ($\lambda_2$-excited) states in the configuration $(bb)(\bar{c}(cc))$, i.e., $M\big(\lambda_1, (cc)(\bar{c}(bb))\big){\approx}M\big(\lambda_1, (bb)(\bar{c}(cc))\big)$, $M\big(\lambda_2, (cc)(\bar{c}(bb))\big){\approx}M\big(\lambda_2, (bb)(\bar{c}(cc))\big)$. Possible mixings among these states are not considered here.

Fig. \ref{fig:lambda} shows the radial and orbital $\lambda_1$- and $\lambda_2$-{\trs} for the pentaquark $P_{cc\bar{c}bb}$ in the configurations $(cc)(\bar{c}(bb))$ and $(bb)(\bar{c}(cc))$. Circles represent the predicted data, calculated using Eqs. (\ref{ppt2q}) and (\ref{pppa2qQ}) or Eqs. (\ref{rtpf}) and (\ref{pppa2qQ}), as listed in Table \ref{tab:lambda}. The black dashed lines represent the complete forms of the $\lambda$-{\trs}, obtained from Eqs. (\ref{ppt2q}) and (\ref{pppa2qQ}) or from Eqs. (\ref{rtpf}) and (\ref{pppa2qQ}). The pink lines represent the fitted formulas, obtained by a linear fit to the calculated data in Table \ref{tab:lambda} and listed in Table \ref{tab:formulas}. The blue dotted lines correspond to the main parts of the full forms, which are also given in Table \ref{tab:formulas}.
Since the diquark $1$ and the triquark are treated as individual constituents without considering their internal substructures, the $\lambda_1$-trajectories are the simplest among the four series of {\rts} (see Table \ref{tab:formulas} and Fig. \ref{fig:lambda}). In Fig. \ref{fig:lambda}, the black dashed lines, pink lines, and blue dotted lines for the $\lambda_1$-trajectories overlap; that is, the compete forms, fitted formulas, and main parts are identical for these trajectories.
Similar to the $\rho$-{\trs}, the explicit forms of the $\lambda_2$-{\trs} are lengthy. Performing a linear fit to the calculated data yields the fitted formulas in Table \ref{tab:formulas}. As shown in Fig. \ref{fig:lambda}, these fitted formulas closely match the complete forms of the $\lambda_2$-trajectories, indicating that the full forms can be well approximated by the simple fitted formulas. Therefore, both the complete forms and the fitted formulas for the $\lambda_2$-trajectories exhibit the behavior $M{\sim{x_{\lambda_2}^{2/3}}}$. However, the main parts for the complete forms of the $\lambda_2$-{\trs} (listed in Table \ref{tab:formulas}) deviate from the full expressions.

\begin{table*}[!phtb]
\caption{Masses of the ground and first excited states of the pentaquark $P_{cc\bar{c}bb}$. The notation in Eq. (\ref{tetnot}) is expressed as $|n_1^{2s_1+1}l_{1j_1},n_2^{2s_2+1}l_{2j_2},N_2^{2s_{3}+1}L_{2J_2},N_1^{2s_4+1}L_{1J_1}\rangle$. $|n_1^{2s_1+1}l_{1j_1},n_2^{2s_2+1}l_{2j_2},N_2L_{2},N_1L_{1}\rangle$ denotes the spin-averaged states. The data are taken from Tables \ref{tab:rho} and \ref{tab:lambda}. ${\dagger}$ denotes that for the first five states, the angular momenta for different structures are all equal to zero, $l_1=l_2=L_1=L_2=0$. ${\ddagger}$ denotes that for the $|1^3s_1, 1^3s_1, 1P, 1S\rangle$ state, $l_1=l_2=L_1=0$, $L_2=1$. ${\ast}$ denotes that for the $|1^3s_1, 1^3s_1, 1S, 1P\rangle$ state, $l_1=l_2=L_2=0$, $L_1=1$.}  \label{tab:sum}
\centering
\begin{tabular}{cccc}
\hline\hline
$J^P$& $|n_1^{2s_1+1}l_{1j_1},n_2^{2s_2+1}l_{2j_2},N_2L_{2},N_1L_{1}\rangle$        & $(cc)(\bar{c}(bb))$ [{\gev}] &  $(bb)(\bar{c}(cc))$ [{\gev}]  \\
\hline
$^{\dagger}$$\left(\frac{1}{2}\right)^{-}$, $\left(\frac{3}{2}\right)^-$, $\left(\frac{5}{2}\right)^-$
 & $|1^3s_1, 1^3s_1, 1S, 1S\rangle$  & 14.13   & 14.10    \\
 & $|2^3s_1, 1^3s_1, 1S, 1S\rangle$  & 14.49   & 14.41  \\
 & $|1^3s_1, 2^3s_1, 1S, 1S\rangle$  & 14.45   & 14.46  \\
 & $|1^3s_1, 1^3s_1, 2S, 1S\rangle$  & 14.45   & 14.43  \\
 & $|1^3s_1, 1^3s_1, 1S, 2S\rangle$  & 14.58   & 14.54  \\
$^{\ddagger}$$\left(\frac{1}{2}\right)^+$, $\left(\frac{3}{2}\right)^+$, $\left(\frac{5}{2}\right)^+$,
$\left(\frac{7}{2}\right)^+$
 & $|1^3s_1, 1^3s_1, 1P, 1S\rangle$  & 14.36   & 14.33  \\
$^{\ast}$$\left(\frac{1}{2}\right)^+$, $\left(\frac{3}{2}\right)^+$, $\left(\frac{5}{2}\right)^+$, $\left(\frac{7}{2}\right)^+$
 & $|1^3s_1, 1^3s_1, 1S, 1P\rangle$  & 14.45   & 14.42  \\
\hline\hline
\end{tabular}
\end{table*}

For ease of comparison with possible experimental results and other theoretical approaches, the masses of the ground and first excited states of the pentaquark $P_{cc\bar{c}bb}$ are summarized in Table \ref{tab:sum}.
For the first five states, the angular momenta for different structures are all zero. In general, the mass of the pentaquark $\left(\frac{1}{2}\right)^-$ state formed by a spin-$\frac{1}{2}$ triquark and a spin-1 diquark is not equal to that of the pentaquark $\left(\frac{1}{2}\right)^-$ state formed by a spin-$\frac{3}{2}$ triquark and a spin-1 diquark, although the mass difference is small. The same applies for the pentaquark $\left(\frac{3}{2}\right)^-$ state.
For the $|1^3s_1, 1^3s_1, 1P, 1S\rangle$ state, the pentaquark $\left(\frac{1}{2}\right)^+$ state can be formed by the spin-1 diquark and either a spin-$\frac{1}{2}$ triquark with $s_3=\frac{1}{2}$, a spin-$\frac{3}{2}$ triquark with $s_3=\frac{1}{2}$, a spin-$\frac{1}{2}$ triquark with $s_3=\frac{3}{2}$, or a spin-$\frac{3}{2}$ triquark with $s_3=\frac{3}{2}$.
For the $|1^3s_1, 1^3s_1, 1S, 1P\rangle$ state, the pentaquark $\left(\frac{1}{2}\right)^+$ state can be formed by the spin-1 diquark and either a spin-$\frac{1}{2}$ triquark with $s_4=\frac{1}{2}$, a spin-$\frac{1}{2}$ triquark with $s_4=\frac{3}{2}$, a spin-$\frac{3}{2}$ triquark with $s_4=\frac{1}{2}$, or a spin-$\frac{3}{2}$ triquark with $s_4=\frac{3}{2}$.
The masses of these penatquark $\left(\frac{1}{2}\right)^+$ states will differ from one another; however, the mass differences among these states will be small. Other states can be discussed in a similar manner.
These mass differences are relativistic corrections, which do not affect the {\rt} behavior. They are not considered here but will be addressed in future work.

\section{Discussions}\label{sec:dis}

To date, there is no evidence of a fully heavy pentaquark. However, many theoretical studies have been conducted on fully heavy pentaquarks \cite{Gordillo:2024blx,Liang:2024met,Gordillo:2023tnz,An:2020jix,An:2022fvs,
Sharma:2025adr,Yan:2021glh,Zhang:2023hmg,Wang:2021xao,Azizi:2025fmx,Zhang:2020vpz,
Yang:2022bfu,Rashmi:2024ako}. These theoretical studies will aid in the experimental search for fully heavy pentaquarks. In this work, we present the $\rho$- and $\lambda$-{\trs} for the fully heavy pentaquark $P_{cc\bar{c}bb}$, and apply the pentaquark Regge trajectory to provide a rough estimate of the masses of $P_{cc\bar{c}bb}$ (see Tables \ref{tab:rho} and \ref{tab:lambda}).
The pentaquark {\rt}, given by Eqs. (\ref{ppt2q}) and (\ref{pppa2qQ}), or alternatively by Eqs. (\ref{rtpf}) and (\ref{pppa2qQ}), can also be applied to fully charmed pentaquarks and other fully heavy pentaquarks. Note that when $\mu_{\lambda}>3.83$ {\gev}, the relations in Eq. (\ref{fitcfxnr}) should be adjusted by fitting to experimental and theoretical data \cite{Xie:2024dfe}.

To our knowledge, this is the first systematic discussion of all four series of {\rts} for fully heavy pentaquarks (see also Ref. \cite{Song:2024bkj}, which cites the results of this work).
Previously, the relation in Eq. (\ref{massform}), the potential in Eq. (\ref{potv}), and the parameter values in Table \ref{tab:parmv} have been successfully applied to discuss the {\rts} for mesons, baryons, diquarks, triquarks, and tetraquarks \cite{Chen:2023djq,Chen:2022flh,
Feng:2023txx,Xie:2024lfo,Xie:2024dfe,Song:2024bkj}. They are now employed to study pentaquark $P_{cc\bar{c}bb}$, yielding results consistent with other theoretical predictions (see Table \ref{tab:masscompc}). This not only demonstrates the universality of the {\rt} relation and parameter values, but also illustrates its predictive capability.

\begin{table}[!phtb]
\caption{Comparison of theoretical predictions for the spin-averaged
masse of the ground state of pentaquark $P_{cc\bar{c}bb}$.}  \label{tab:masscompc}
\centering
\begin{tabular*}{0.46\textwidth}{@{\extracolsep{\fill}}ccc@{}}
\hline\hline
 & Mass [{\gev}] & Method\\
\hline
$\{cc\}(\bar{c}\{bb\})$ & 14.13  & {\rt} \\
$\{bb\}(\bar{c}\{cc\})$ & 14.10  & {\rt}\\
$\{bc\}(\bar{c}\{bc\})$ & 14.13  & {\rt}\\
Ref. \cite{Rashmi:2024ako} &  15.21 &  Effective quark mass  \\
                       &         & and screened charge scheme \\
Ref. \cite{Liang:2024met} &  14.62  & Nonrelativistic potential quark model\\
Ref. \cite{Gordillo:2024blx}  & 14.30 & Schr\"{o}dinger equation\\
Ref. \cite{An:2022fvs}        &14.57  & Constituent quark model\\
Ref. \cite{Zhang:2023hmg}     & 14.86 & MIT bag model \\
\hline\hline
\end{tabular*}
\end{table}

One merit of the Regge trajectory approach is its simple analytical form. However, the complete forms of the $\rho_1$-, $\rho_2$-, and $\lambda_2$-trajectories for pentaquarks are quite lengthy and cumbersome (see Eqs. (\ref{fulla}) and (\ref{fullb})). We fit the calculated data to obtain simplified relations, as listed in Table \ref{tab:formulas}. Because the used data are calculated by using the complete forms of the {\rts}, the lengthy complete forms can be well approximated by the simple fitted relations, and they have the same behaviors (see Figs. \ref{fig:rho} and \ref{fig:lambda}).

It is noteworthy that the main parts of the complete forms and the fitted formulas share the same functional form, $M=m_R+\beta(x+c_0)^{2/3}$. However, the values of the parameters $m_R$ and $\beta$ differ considerably.
For example, for the main part of the complete form of the $\rho_2$-{\tr}, $m_R=13.8296$ and $\beta=0.332799$; these values are readily computed using Eqs. (\ref{ppt2q}) and (\ref{pppa2qQ}), and clearly depend on the constituents' masses and the string tension. By contrast, for the fitted $\rho_2$-{\tr}, $m_R=14.1131$ and $\beta=0.331141$; here, the dependence on the constituents' masses and string tension is not readily apparent and becomes more complicated. This indicates that, unlike in the meson case, the dependence of $m_R$ and $\beta$ on the constituents' masses and string tension is no longer obvious or direct when we construct simply fitted $\rho_1$-, $\rho_2$-, and $\lambda_2$-{\tr} formulas for pentaquarks.
As discussed above and in Sec. \ref{sec:rts}, the $\rho_1$, $\rho_2$, and $\lambda_2$ {\rts} for fully heavy pentaquarks cannot be obtained by merely mimicking the meson {\rts}; instead, they should be constructed based on the actual structure and substructure of the pentaquark.
Otherwise, the $\rho_1$-, $\rho_2$-, and $\lambda_2$-trajectories must rely solely on fitting existing theoretical or future experimental data. Consequently, the fundamental relationship between the slopes of the obtained trajectories and constituents' masses and string tension will become unobvious and complicated. The predictive power of the Regge trajectories would be compromised.

The Regge trajectories assume different forms and exhibit distinct behaviors across different energy regions \cite{Chen:2022flh,Chen:2021kfw}.
The diquark $(bb)$, $(cc)$ and the triquark $(\bar{c}(cc))$, $(\bar{c}(bb))$ are all the heavy-heavy systems; therefore, the main parts of the $\rho_1$-, $\rho_2$-, $\lambda_1$-, and $\lambda_2$-trajectories all follow the behavior $M{\sim}x^{2/3}$, where  $x=x_{\rho_1},\,x_{\rho_2},\,x_{\lambda_1},\,x_{\lambda_2}$ (see Eq. (\ref{rtpf})).
Fitting the calculated data shows that the complete forms of the Regge trajectories also exhibit the same behavior as their main parts, namely $M{\sim}x^{2/3}$, see Table \ref{tab:formulas}.

For potential models, both the dynamic equation and the potential determine the form of the {\rts} \cite{Chen:2022flh,Chen:2021kfw,Chen:2018bbr}. Constituents are classified as light or heavy based on their masses, which correspond to ultrarelativistic or nonrelativistic dynamical equations, leading to different Regge trajectories \cite{Chen:2022flh,Chen:2021kfw}.
At large $r$, the color Coulomb potential is neglected, while the confining potential plays a dominant role in potential (\ref{potv}), which affects the form of {\rts} \cite{brau:04bs,Sonnenschein:2018fph}. The parameters $\sigma$ and $C$ in Eq. (\ref{mrcc}) do not affect the forms of the Regge trajectories but do affect their values; for example, see Eqs. (\ref{ppt2q}) and (\ref{pppa2qQ}).
For fully heavy pentaquarks, all four series of {\rts} are concave downward in the $(M,\,x)$ plane when the confining potential is linear. It can also be easily shown that these {\trs} are concave downward in the $(M^2,\,x)$ plane under the same conditions.
For light systems, such as light mesons, the Cornell potential has the same functional form and, when the masses of the light constituents are neglected, results in linear Regge trajectories in the $(M^2,\,x)$ plane \cite{Chen:2022flh,Chen:2021kfw,Lucha:1991vn}. When the masses of the light constituents are taken into account, the Regge trajectories for light systems also become concave in the $(M^2,\,x)$ plane \cite{Wilczek:2004im,Chen:2023web}. It is thus expected that the Regge trajectories for light, singly heavy, doubly heavy, triply heavy, quadruply heavy, and fully heavy pentaquarks are all concave downward in the $(M^2,\,x)$ plane when the confining potential is linear and the masses of the light quarks are considered.

\section{Conclusions}\label{sec:conc}
In this work, we propose {\rt} relations for the fully heavy pentaquark $P_{cc\bar{c}bb}$, employing both diquark and triquark {\rts}. By employing these new relations, we present four series of Regge trajectories for the pentaquark $P_{cc\bar{c}bb}$ in the configurations $(cc)(\bar{c}(bb))$ and $(bb)(\bar{c}(cc))$: the $\rho_1$-, $\rho_2$-, $\lambda_1$-, and $\lambda_2$-trajectories. The masses of the $\rho_1$-, $\rho_2$-, $\lambda_1$-, and $\lambda_2$-excited states are roughly estimated.

The complete forms of the Regge trajectories for pentaquark $P_{cc\bar{c}bb}$ are often lengthy and cumbersome. Except for the $\lambda_1$-{\trs}, the $\rho_1$-, $\rho_2$-, and $\lambda_2$-{\trs} for the fully heavy pentaquarks cannot be obtained by simply mimicking the meson {\rts} but should instead be constructed based on pentaquark's structure and substructure.
Otherwise, the $\rho_1$-, $\rho_2$-, and $\lambda_2$-trajectories must rely solely on fitting existing theoretical or future experimental data. The fundamental relationship between the slopes of the obtained trajectories and constituents' masses and string tension will become unobvious. The predictive power of the Regge trajectories would be compromised.

We show that the lengthy complete forms of the $\rho_1$-, $\rho_2$-, and $\lambda_2$-trajectories can be well approximated by the simple fitted formulas. All four series of Regge trajectories for the pentaquark $P_{cc\bar{c}bb}$ exhibit a behavior of $M{\sim}x^{2/3}$, where  $x=n_{r_1},n_{r_2},l_1,l_2,N_{r_1},N_{r_2},L_1,L_2$.
Moreover, all four series of trajectories are concave downward in the $(M^2,\,x)$ plane.

\section*{Acknowledgments}
We are very grateful to the anonymous referees for the valuable comments and suggestions.

\appendix
\section{List of the {\rt} relations}\label{sec:appa}
The concrete forms of the {\rts} for the pentaquark are not as same simple as those for mesons. Although Eqs. (\ref{ppt2q}) and (\ref{pppa2qQ}) or Eqs. (\ref{rtpf}) and (\ref{pppa2qQ}) are simple, their final forms can be rather tedious, because pentaquarks have substructures and the slope depends on the masses of the constituents.

From Eqs. (\ref{ppt2q}) and (\ref{pppa2qQ}) or Eqs. (\ref{rtpf}) and (\ref{pppa2qQ}), we can easily obtain the complete forms of the {\rts} for the fully heavy pentaquark $P_{cc\bar{c}bb}$. The resulting expressions are rather long and tedious. As an example, we list only the radial $\rho_1$- and $\rho_2$-{\trs} for the configuration $(cc)(\bar{c}(bb))$. The $\rho_1$-{\tr} for the configuration $(cc)(\bar{c}(bb))$ is given by
\begin{widetext}
\begin{align}\label{fulla}
M=&13.8255+0.487766(0.205+n_{r_1})^{2/3}
+0.318109 \left(\frac{14.1255+0.487766 (0.205+n_{r_1})^{2/3}}{2.95+0.487766 (0.205+n_{r_1})^{2/3}}\right)^{1/3}\nonumber\\
&\times\left(0.334-\frac{0.97227 \left(2.95+0.487766 (0.205+n_{r_1})^{2/3}\right)}{14.1255+0.487766 (0.205+n_{r_1})^{2/3}}\right)^{2/3}
\left(1.008+\frac{0.0894041 \left(2.95+0.487766 (0.205+n_{r_1})^{2/3}\right)}{14.1255+0.487766 (0.205+n_{r_1})^{2/3}}\right).
\end{align}
The $\rho_2$-{\tr} for the configuration $(cc)(\bar{c}(bb))$ reads
\begin{align}\label{fullb}
M=&13.8296+0.332799 (0.01+n_{r_2})^{2/3}\nonumber\\
&+0.38714 \left(\frac{11.16+0.332799 (0.01+n_{r_2})^{2/3}}{9.61+0.332799 (0.01+n_{r_2})^{2/3}}\right)^{1/3}\left(0.334-\frac{0.13485 (9.61+0.332799 (0.01+n_{r_2})^{2/3})}{11.16+0.332799 (0.01+n_{r_2})^{2/3}}\right)^{2/3} \nonumber \\
&\times\left(1.008+\frac{0.0124 (9.61+0.332799 (0.01+n_{r_2})^{2/3})}{11.16+0.332799 (0.01+n_{r_2})^{2/3}}\right)+
0.486744 \Big(\Big(14.1296+0.332799 (0.01+n_{r_2})^{2/3}\nonumber\\
&+0.38714 \Big(\frac{11.16+0.332799 (0.01+n_{r_2})^{2/3}}{9.61+0.332799 (0.01+n_{r_2})^{2/3}}\Big)^{1/3} \Big(0.334-\frac{0.13485 (9.61+0.332799 (0.01+n_{r_2})^{2/3})}{11.16+0.332799 (0.01+n_{r_2})^{2/3}}\Big)^{2/3}\nonumber \\
 &\Big(1.008+\frac{0.0124 (9.61+0.332799 (0.01+n_{r_2})^{2/3})}{11.16+0.332799 (0.01+n_{r_2})^{2/3}}\Big)\Big)\Big/\Big(11.01+0.332799 (0.01+n_{r_2})^{2/3}\nonumber\\
 &+0.38714 \Big(\frac{11.16+0.332799 (0.01+n_{r_2})^{2/3}}{9.61+0.332799 (0.01+n_{r_2})^{2/3}}\Big)^{1/3}\Big(0.334-\frac{0.13485 (9.61+0.332799 (0.01+n_{r_2})^{2/3})}{11.16+0.332799 (0.01+n_{r_2})^{2/3}}\Big)^{2/3} \nonumber\\
 &\Big(1.008+\frac{0.0124 (9.61+0.332799 (0.01+n_{r_2})^{2/3})}{11.16+0.332799 (0.01+n_{r_2})^{2/3}}\Big)\Big)\Big)^{1/3} \Big(0.334-\Big(0.271404 \Big(11.01+0.332799 (0.01+n_{r_2})^{2/3}\nonumber\\
 &+0.38714 \Big(\frac{11.16+0.332799 (0.01+n_{r_2})^{2/3}}{9.61+0.332799 (0.01+n_{r_2})^{2/3}}\Big)^{1/3} \Big(0.334-\frac{0.13485 (9.61+0.332799 (0.01+n_{r_2})^{2/3})}{11.16+0.332799 (0.01+n_{r_2})^{2/3}}\Big)^{2/3}\nonumber\\
  &\Big(1.008+\frac{0.0124 (9.61+0.332799 (0.01+n_{r_2})^{2/3})}{11.16+0.332799 (0.01+n_{r_2})^{2/3}}\Big)\Big)\Big)\Big/\Big(14.1296+0.332799 (0.01+n_{r_2})^{2/3}\nonumber \\
  &+0.38714 \Big(\frac{11.16+0.332799 (0.01+n_{r_2})^{2/3}}{9.61+0.332799 (0.01+n_{r_2})^{2/3}}\Big)^{1/3} \Big(0.334-\frac{0.13485 (9.61+0.332799 (0.01+n_{r_2})^{2/3})}{11.16+0.332799 (0.01+n_{r_2})^{2/3}}\Big)^{2/3}\nonumber 
\end{align}
\begin{align}   
   &\Big(1.008+\frac{0.0124 (9.61+0.332799 (0.01+n_{r_2})^{2/3})}{11.16+0.332799 (0.01+n_{r_2})^{2/3}}\Big)\Big)\Big)^{2/3} \Big(1.008+\Big(0.0249567 \Big(11.01+0.332799 (0.01+n_{r_2})^{2/3}\nonumber \\
   &+0.38714 \Big(\frac{11.16+0.332799 (0.01+n_{r_2})^{2/3}}{9.61+0.332799 (0.01+n_{r_2})^{2/3}}\Big)^{1/3} \Big(0.334-\frac{0.13485 (9.61+0.332799 (0.01+n_{r_2})^{2/3})}{11.16+0.332799 (0.01+n_{r_2})^{2/3}}\Big)^{2/3}\nonumber\\
    &\Big(1.008+\frac{0.0124 (9.61+0.332799 (0.01+n_{r_2})^{2/3})}{11.16+0.332799 (0.01+n_{r_2})^{2/3}}\Big)\Big)\Big)\Big/\Big(14.1296+0.332799 (0.01+n_{r_2})^{2/3}\nonumber\\
    &+0.38714 \Big(\frac{11.16+0.332799 (0.01+n_{r_2})^{2/3}}{9.61+0.332799 (0.01+n_{r_2})^{2/3}}\Big)^{1/3} \Big(0.334-\frac{0.13485 (9.61+0.332799 (0.01+n_{r_2})^{2/3})}{11.16+0.332799 (0.01+n_{r_2})^{2/3}}\Big)^{2/3} \nonumber\\
    & \Big(1.008+\frac{0.0124 (9.61+0.332799 (0.01+n_{r_2})^{2/3})}{11.16+0.332799 (0.01+n_{r_2})^{2/3}}\Big)\Big)\Big).
\end{align}
\end{widetext}

The $\rho_1$- and $\rho_2$-excited masses for the configuration $(cc)(\bar{c}(bb))$ are calculated using Eqs. (\ref{fulla}) and (\ref{fullb}), respectively, and are listed in Table \ref{tab:rho}. By linearly fitting the calculated masses in the $(M, (c_0+x)^{2/3})$ plane, we obtain the fitted formulas:
\bea\label{fra}
M=13.9661+ 0.465794(0.205+n_{r_1})^{2/3}
\eea
and
\bea\label{frb}
M=14.1131+ 0.331141(0.01+n_{r_2})^{2/3},
\eea
respectively. The complete forms of different Regge trajectories exhibit varying complexity [see Eqs. (\ref{fulla}) and (\ref{fullb})]. However, their fitted counterparts, as shown in Eqs. (\ref{fra}) and (\ref{frb}), are comparatively simple.
Moreover, $m_R$ and slope for the fitted {\rts} [Eqs. (\ref{fra}) and (\ref{frb})] are different from those for the main parts of the complete forms of the {\rts} [see Eqs. (\ref{fulla}) and (\ref{fullb}) or Table \ref{tab:formulas}.]. For the $\rho_1$-{\tr}, $m_R=13.9661$ for the fitted form, while it is $13.8255$ for the main part. The values of $\beta_{n_{r_1}}$ are $0.465794$ for the fitted $\rho_1$-{\tr} and $0.487766$ for the main part of the complete form, respectively.

The explicit forms of the {\rts} calculated from Eqs. (\ref{ppt2q}) and (\ref{pppa2qQ}) or from Eqs. (\ref{rtpf}) and (\ref{pppa2qQ}) are often rather tedious. Here, we present only the main parts of the complete forms of the Regge trajectories, as well as the fitted formulas obtained by fitting the calculated results (see Table \ref{tab:formulas}).

\begin{table*}[!phtb]
\caption{The fitted formulas and the main parts of the $\rho$- and $\lambda$-{\rts} for pentaquark $P_{cc\bar{c}bb}$ in the configurations $(cc)(\bar{c}bb)$ and $(bb)(\bar{c}cc)$. $({\rm fit})$ denotes the formula obtained by fitting the calculated results. $({\rm main})$ indicates the main part of the complete forms of the {\rts}, derived from Eqs. (\ref{ppt2q}) and (\ref{pppa2qQ}).}  \label{tab:formulas}
\centering
\begin{tabular*}{1.0\textwidth}{@{\extracolsep{\fill}}ccc@{}}
\hline\hline
        & $(cc)(\bar{c}(bb))$ [{\gev}] &  $(bb)(\bar{c}(cc))$ [{\gev}]  \\
\hline
 $\rho_1$  & $M=13.9661+ 0.465794(0.205+n_{r_1})^{2/3}$ (fit)   &
          $M=14.0817+ 0.329997 (0.01+n_{r_1})^{2/3}$  (fit)     \\
 & $M=13.8255+0.487766 (0.205+n_{r_1})^{2/3}$ (main)   &
 $M=14.0053+0.332799 (0.01+n_{r_1})^{2/3}$ (main)    \\
   & $M=13.9663+ 0.366186 (0.337+l_1)^{2/3}$ (fit) &
   $M=14.0817+ 0.259577(0.001+l_1)^{2/3}$ (fit)\\
   & $M=13.8255+0.383712 (0.337+l_1)^{2/3}$ (main) &
   $M=14.0053+0.261804 (0.001+l_1)^{2/3}$ (main)\\
 $\rho_2$  & $M=14.1131+ 0.331141(0.01+n_{r_2})^{2/3}$ (fit)   &
       $M=13.9352+ 0.465439 (0.205+n_{r_2})^{2/3}$ (fit)  \\
           & $M=13.8296+0.332799 (0.01+n_{r_2})^{2/3}$ (main) &
           $M=13.6754+0.487766 (0.205+n_{r_2})^{2/3}$ (main)\\
           & $M=14.1131+ 0.26048 (0.001+l_2)^{2/3}$ (fit) &
           $M=13.9351+ 0.366203(0.337+l_2)^{2/3}$ (fit)\\
           & $M=13.8296+0.261804 (0.001+l_2)^{2/3}$ (main) &
           $M=13.6754+0.383712 (0.337+l_2)^(2/3)$ (main)\\
 $\lambda_1$  & $M=13.9951+ 0.542908(0.121824+N_{r_1})^{2/3}$ (fit)   &    $M=14.0207+0.500992 (0.0594404+N_{r_1})^{2/3}$ (fit)\\
              & $M=13.9951+0.542908 (0.121824+N_{r_1})^{2/3}$ (main) &  $M=14.0207+0.500992 (0.0594404+N_{r_1})^{2/3}$ (main)            \\
              & $M=13.9951+ 0.407734(0.196129+L_1)^{2/3}$ (fit) &
              $M=14.0207+0.377205 (0.0950241+L_1)^{2/3}$ (fit) \\
               & $M=13.9951+ 0.407734(0.196129+L_1)^{2/3}$ (main) &
               $M=14.0207+0.377205 (0.0950241+L_1)^{2/3}$ (main)\\
 $\lambda_2$  & $M=13.979+ 0.413004(0.217853+N_{r_2})^{2/3}$ (fit)  &    $M=13.9281+ 0.433149(0.243911+N_{r_2})^{2/3}$ (fit)\\
              &  $M=13.845+0.414497 (0.217853+N_{r_2})^{2/3}$ (main)  &  $M=13.845+0.450067 (0.243911+N_{r_2})^{2/3}$ (main) \\
              & $M=13.979+ 0.308932(0.351762+L_2)^{2/3}$ (fit) &
              $M=13.9279+ 0.32383 (0.393994+L_2)^{2/3}$ (fit)\\
              & $M=13.845+0.310068 (0.351762+L_2)^{2/3}$ (main) &
             $M=13.845+0.336311 (0.393994+L_2)^{2/3}$ (main)\\
\hline\hline
\end{tabular*}
\end{table*}

\noindent{\bf Data Availability Statement.} This article has no associated data or the data will not be deposited.

\noindent{\bf Code Availability Statement.} This article has no associated code or the code will not be deposited.

\noindent{\bf Open Access.} This article is distributed under the terms of the Creative Commons Attribution License (CC-BY4.0), which permits any use, distribution and reproduction in any medium, provided the original author(s) and source are credited.

\end{document}